\documentclass{article}
\usepackage{lineno,hyperref}
\usepackage[T1]{fontenc}
\usepackage{graphicx,amsmath}
\usepackage{soul,color}
\usepackage{siunitx}
\usepackage[sorting=none,citestyle=numeric-comp]{biblatex}
\bibliography{refs}
\usepackage{lineno}
\usepackage[bottom=2cm, top=2cm, right=2cm, left=2.5cm]{geometry}
\def \art_title{A Predictive Theory of Electrochemical Ostwald Ripening for Electrodeposited Lithium Metal}
\title{\art_title}

\def \authors{Hanning Zhang$^\dagger$, Oleg V. Yazyev$^\dagger$, and Ruslan Yamaletdinov$^{\dagger,*}$}
\author{\authors}

\date{
$^\dagger$Institute of Physics, Ecole Polytechnique Fédérale de Lausanne (EPFL), CH-1015 Lausanne, Switzerland\\
$^*$Corresponding author: \href{mailto:ruslan.yamaletdinov@epfl.ch}{ruslan.yamaletdinov@epfl.ch}\\
}

\begin{document}

\maketitle


\begin{abstract}

Electrode morphology critically determines the stability and efficiency of lithium metal anodes, yet no predictive framework has explained how measurable parameters control deposition. Here we introduce the first theoretical model of electrochemical Ostwald ripening, capturing the competition between electroplating and surface-energy–driven redistribution and identifying it as the governing process behind morphology evolution in the non-dendritic regime. The framework explicitly incorporates SEI resistance, electrolyte conductivity, electrode wettability, and current density revealing the transition from 2D SEI-limited to 3D electrolyte-limited growth. The model yields analytical expressions for nucleus size, density and distribution that quantitatively reproduce independent experimental results and establishes a direct link between plating conditions, morphology, and Coulombic efficiency. By providing experimentally accessible relationships between key parameters and deposition outcomes, the framework enables predictive understanding of lithium plating and provides a broadly applicable basis for controlling electrodeposition morphology across diverse electrochemical systems.

\end{abstract}

\section*{Introduction}

The morphology of electrodeposited lithium is a critical factor in the Coulombic efficiency (CE) and lifespan of lithium metal batteries~\cite{C9CS00883G}. Irregular deposition accelerates solid electrolyte interphase (SEI) degradation, increases electrolyte consumption, and raises the risk of short circuits—all of which contribute to battery failure~\cite{C9CS00883G,JETYBAYEVA2023232914,Monroe_2003}. Conversely, smooth and compact lithium deposits improve cycling stability by reducing side reactions and ensuring uniform lithium utilization.
Beyond batteries, metal deposition morphology is also critical in general galvanic coatings, where surface roughness and deposit uniformity influence properties such as corrosion resistance, adhesion strength, and electrical conductivity~\cite{Schneider2017}. 


Metal deposition can be broadly divided into dendritic and non-dendritic regimes. In the dendritic regime, rapid ion depletion near the electrode surface drives growth into needle-like dendrites~\cite{Barton1962,Sandstime}, posing immediate safety risks. The non-dendritic regime, by contrast, corresponds to practical operating conditions and is the main focus of this work. Here, deposition proceeds more gradually and is generally described as a two-stage process: during nucleation, new nuclei form and their initial distribution is established; during subsequent growth, existing nuclei expand while little additional nucleation occurs~\cite{Stark_2013,Feng2023}.

Classical nucleation theory predicts that large nuclei are statistically suppressed due to the $\sim r^{-2}$ dependence of the Zeldovich factor~\cite{Russell1969}, favoring the formation of small initial clusters under low overpotential $\eta$. The critical nucleus size for lithium is approximately $r^* \sim \frac{2\sigma V_m}{RT}$ (around 18 nm~\cite{Ely2013}), where $\sigma$ is the metal–electrolyte surface energy. Experimentally observed nuclei, however, are often much larger, indicating that growth processes dominate shortly after nucleation under practical conditions~\cite{Feng2023}, highlighting the central role of the growth phase in determining the final deposition morphology. 

\begin{figure}[!h]
     \centering
     \includegraphics[width=0.5\linewidth]{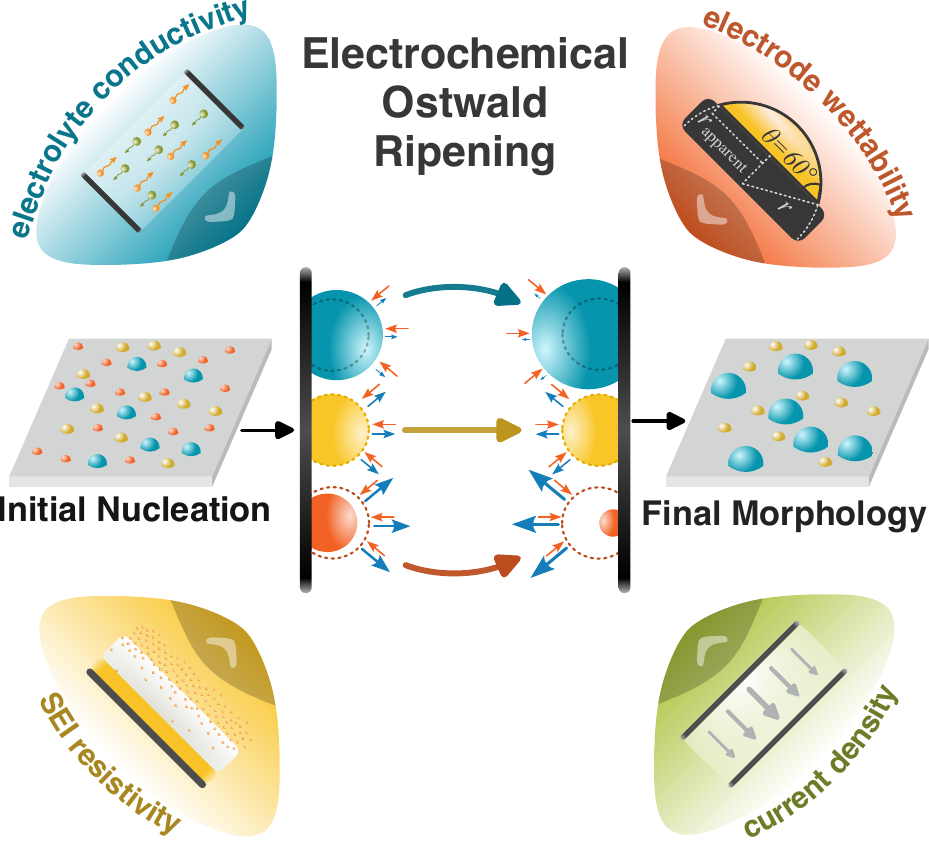}
     \caption{\textbf{Ostwald ripening framework mediating the influence of key factors on lithium morphology.} The central sequence illustrates the evolution from an initially broad distribution of small nuclei to a final morphology dominated by larger ones, highlighting the growth of larger nuclei at the expense of smaller ones. 
}
     \label{fig_1}
 \end{figure}

In classical models the growth is assumed to proceed without dissolution of nuclei, which leads at constant current to a linear increase in volume, $r^3 \sim it$. This simplified picture omits redistribution processes driven by surface energy minimization, commonly known as Ostwald ripening, in which larger nuclei grow at the expense of smaller ones~\cite{Lifshitz1961,Wagner1961} (See Fig~\ref{fig_1}). However, ripening is crucial for two reasons. First, it accounts for the experimentally observed decrease in nucleation density during the early stages of lithium deposition~\cite{Pei2017,Yan2019} and reflects similar observations in other metal systems, including sodium-ion batteries~\cite{Redmond2005,Geng2022,Hwang2023}, where redistribution promotes smoother, denser deposits. Second, the interplay between electroplating and redistribution current proves necessary to explain the roles of key parameters such as SEI and electrolyte resistivity, metal–electrode interfacial energy, and current density in shaping the final morphology. This addresses an important theoretical gap; although these factors are experimentally known to influence lithium electrodeposition, with lower values generally improving cell life~\cite{Ye2022,anie.201702099,Kravchyk2024,Wang2019,Que2023}, the mechanisms by which they exert their effect have remained elusive.

A key implication of deposition morphology is its impact on Coulombic efficiency and battery lifespan~\cite{Hobold2021,C9CS00883G}, as a higher surface-to-volume ratio of the lithium deposit accelerates parasitic processes. The morphology at the point of full electrode coverage is particularly decisive, as it defines the subsequent evolution of lithium growth. A simple illustration is provided by the SEI-passivated column model~\cite{D0TA06020H}, where once the electrode surface is fully covered, further lithium growth proceeds only in the vertical direction, with columns isolated by the SEI retaining their initial diameters at coverage point. Even when the morphology becomes more complex—for example, when softer SEI allows partial fusion of neighboring columns and the formation of voids filled with trapped SEI and electrolyte~\cite{BOARETTO2021229919}—the initial state still fully determines the subsequent evolution.

In this study, we derive a theoretical framework that explicitly incorporates electrochemical Ostwald ripening into the description of lithium electrodeposition, demonstrating its previously unrecognized yet decisive role in determining plating morphology in the non-dendritic regime. The model quantifies how deposition conditions and cell parameters—such as SEI and electrolyte resistances, surface energy and plating current — govern the evolution of deposit structure prior to full electrode surface coverage, while the predicted nuclei sizes at this critical point are shown to correlate with CE. Theoretical results closely agree with experimental observations across multiple datasets.
 Although developed for lithium metal anodes, the framework is broadly applicable to other metal electrodeposition processes, providing a unified basis for understanding and controlling metal morphology across diverse technologies.

\section*{Results}
\subsection*{General system description}
In the classical  Ostwald ripening theory of Wagner, Lifshitz, and Slyozov (WLS)\cite{Wagner1961,Lifshitz1961}, the governing equations are derived directly from the diffusion equation. For electrochemical systems, however, this approach is less practical\cite{Schroder2006}, as it introduces redundant parameters—particularly for the SEI—that are difficult to measure experimentally. Instead, we adopt a Barton-type formulation\cite{Barton1962}, which can be directly related to the WLS theory but relies only on experimentally accessible material constants.
In our framework, the total overpotentials is expressed as the sum of electrodic, diffusive, and surface energy contributions~\cite{Barton1962}. In the low-current density regime, the electrodic and SEI overpotentials can be linearized and expressed using Ohm's law~\cite{Barton1962,Schroder2006, Hess2017}, allowing the total overpotential to be written as:  

\begin{equation}
    \label{eq_Barton}
    \begin{aligned} 
    \eta &= \frac{i_n}{i_0} \frac{RT}{F} + i_n R_{SEI} S_{tot}  +  \frac{i_nrRT}{D C_\infty F^2} + \frac{2\sigma  V_m}{Fr} .
    \end{aligned}
\end{equation}
Here $i_n$ is the current density at nucleus surface, $i_0$ is the exchange current density, and $F$ is the Faraday constant. $R_{SEI}$ is the SEI resistance, and $S_{tot}$ is the total SEI surface area.  $D$ and $C_\infty$ are the lithium ion diffusivity and bulk lithium ion concentration of the electrolyte, respectively. The final term, $\frac{2\sigma  V_m}{Fr}$, accounts for the additional overpotential contribution from the surface energy of the lithium nuclei.  

Since the exchange current density $i_0$ is relatively high, with characteristic values on the order of $10^2$–$10^3$ A/m$^2$\cite{Amin2017, Que2023} while the SEI  term can vary between $10^{-1}$ and $10^3$ $\Omega$ cm$^2$\cite{Wang2019, Hobold2023,Solchenbach2021}, the electrodic overpotential is significant only when the SEI is highly conductive. For simplicity, we combine these two terms into a single corrected SEI resistivity, defined as $\Re/\kappa=\frac{RT}{i_0F} +  R_{SEI} S_{tot}$, where $\kappa$ is a free parameter with units of surface conductance ($\Omega^{-1}$m$^{-2}$) introduced solely for notational convenience.

The current density can be related to the rate of change of the nucleus radius as: $i_n = \frac{3 F v}{ V_m s} \frac{dr}{dt}$, where  $v=\frac{\pi}{3}(2+\cos\theta)(1-\cos\theta)^2$ and $s=2\pi (1-\cos\theta)$ are the volumetric and surface shape factors for a hemispherical shape \footnote{\label{fn2} For very small clusters (on the order of $\sim$10 nm), faceted geometries such as rhombic dodecahedra are more thermodynamically preferable ~\cite{FU201879}. Yet as the nuclei grow, the energetic advantage of these faceted shapes over simpler geometries diminishes\cite{Schroder2006}, and their morphology approaches that of truncated spheres defined by the metal–substrate contact angle $\theta$, which reflects the balance of interfacial energies via Young’s equation.}
(see Sec.\ref{sec_units}). Substituting the dimensionless variables $\rho= \frac{RT}{2\sigma V_m}r$, $\rho_s=\frac{RT}{F\eta}$, $\tau=\frac{\kappa(RT)^2}{6 \sigma F^2} \frac{s}{v} t$, and $\omega=\frac{2\sigma V_m\kappa}{D C_\infty F^2 }$ into Eq. \ref{eq_Barton} we obtain:
\begin{equation}
\label{eq_d_rho_d_tau}
v_\rho=\frac{d \rho}{d\tau}= \frac{1}{\Re+ \omega \rho}\left(\frac{1}{\rho_s}-\frac{1}{\rho}\right)
\end{equation}

Eq.~\ref{eq_d_rho_d_tau} describes the dynamics of the nucleus size evolution. Nuclei smaller than the instantaneous stationary radius $\rho_s(\tau)$, indicated as yellow nuclei in Fig.~\ref{fig_1}, dissolve, while nuclei bigger than $\rho_s$ grow. Importantly, the stationary radius $\rho_s(\tau)$ itself is a growing function with time, such that increasingly bigger nuclei start being dissolved as the system evolves. An interesting consequence of Eq.~\ref{eq_d_rho_d_tau} is the emergence of mixed-dimensional behavior in nuclei growth during Ostwald ripening. At early stages (or under high SEI resistivity conditions, $\Re\gg \omega \rho$), the growth dynamics are governed primarily by SEI resistivity, and the system exhibits an effective dimensionality of $n = 2$. In contrast, at later stages (or under low SEI resistivity, $\omega \rho \gg \Re$), the system transitions to behavior consistent with the $n = 3$ dimensional Ostwald ripening equation of motion. Both limits, by choosing the appropriate $\kappa$ such that $\Re = 1$ in the 2 dimensional and $\omega =1$ in the 3 dimensional case, can be unified into a generalized form 

\begin{equation}
\label{eq_d_rho_d_tau_n}
v_\rho=\frac{d\rho}{d\tau}=\frac{1}{\rho^{{n-1}}}\left(\frac{\rho}{\rho_s}-1\right).
\end{equation} Following Lifshitz-Slyozov~\cite{Lifshitz1961}, we further introduce the nuclei size distribution function $f(\rho,\tau)$, which fulfills the continuity equation
\begin{equation}
\label{eq_f_cont}
\frac{d f}{d\tau}+\frac{d}{d \rho}\left(f v_\rho\right)=0
\end{equation}

In the constant current regime, mass conservation is expressed as
\begin{equation}
\label{eq_consv_of_mat_general}
    j=\frac{1}{\tau}\int_0^\infty f(\rho,\tau)\rho^3d\rho=3\int_0^\infty f(\rho,\tau)v_\rho \rho^2d\rho, 
\end{equation}
where $j =  \frac{ F i}{\kappa RT}\frac{3}{s}$ is the dimensionless ion flow. Notably, it is the use of the constant-current assumption in Eq.~\ref{eq_consv_of_mat_general}, replacing the constant-volume restriction employed in previous works, that gives rise to new phenomena explored in this paper (see \ref{sec_sol} for differences).

\subsection*{Asymptotic limits}
\begin{figure}[!h]
 \begin{center}
    \includegraphics[width=0.95\linewidth]{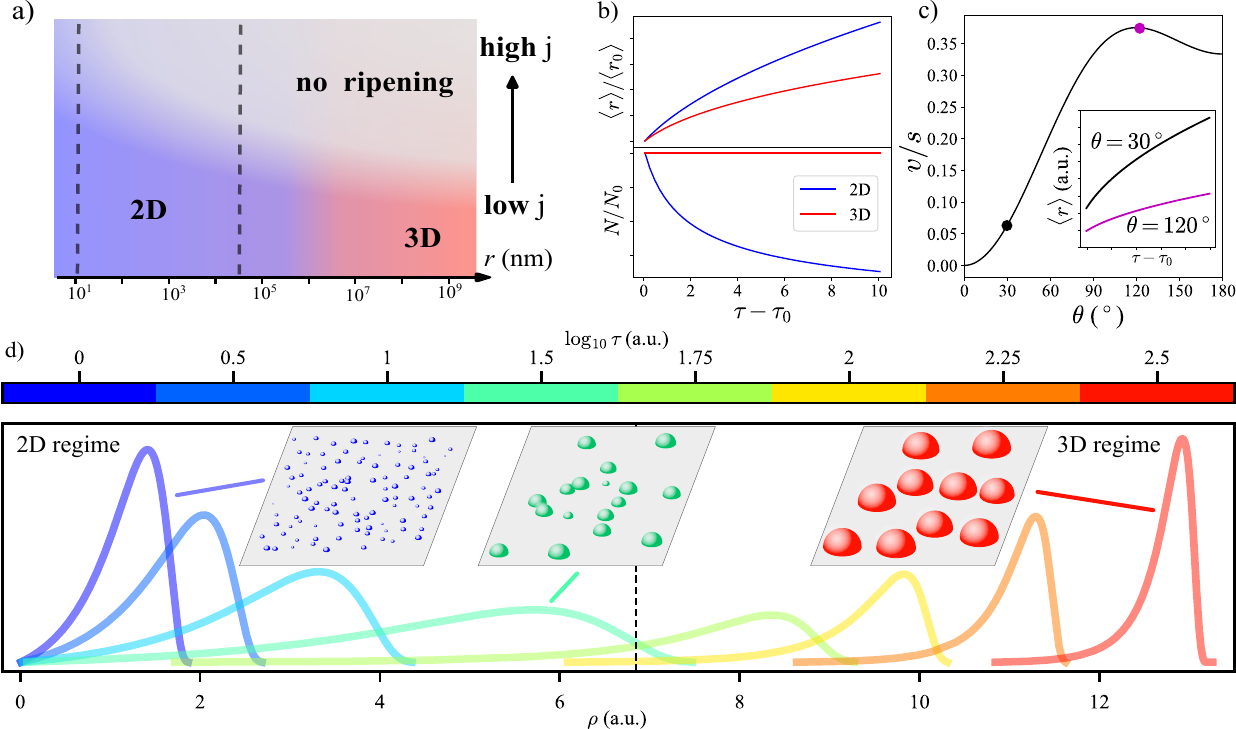}
     
     \caption{\textbf{Comparison of 2D and 3D behavior in constant-current electrochemical Ostwald ripening.} (a) Phase diagram showing the 2D/3D and no ripening regimes. The left and right dotted line mark approximate nuclei sizes at nucleation and at full coverage. 
     (b) Time  evolution of average nucleus size ($\langle r \rangle$, top), and nucleation density ($N$, bottom) in both 2D and 3D regimes. (c) Shape parameter ($v/s$) as a function of contact angle ($\theta$). Inset: Comparison of $\langle r \rangle$ (in units of ${V_m}/{F} \sqrt{{\sigma R_{SEI}S_{tot}}}$ from Eq.~\ref{eq_r_2D}) for two different contact angles. (d) Time evolution of the nucleus size distribution: initially following the $f^\infty_{2D}$ profile, the system gradually transitions to a 3D regime with a narrowing distribution. Different time steps are color-coded; insets illustrate the evolving surface coverage.}
     \label{fig_2}
 \end{center}
 \end{figure}

The solutions of the systems of equation ~\ref{eq_d_rho_d_tau}–\ref{eq_consv_of_mat_general} are discussed in three relevant limiting cases (Fig.\ref{fig_2}(a)). Firstly, for high current densities $j$ or high total charge-transfer resistance, the term $\frac{1}{\rho_s}$ becomes dominant in the nucleus growth $v_\rho$ and ripening becomes negligible. In this regime, provided dendritic formation conditions have not been reached, nucleus growth proceeds conserving the nucleus number density. As a result, the size distribution $f(r,t)$ evolves primarily through a shift of its center—moving toward larger radii, while its shape is approximately preserved\cite{Bartels1991} (see Tab.\ref{tab_1}, Sec.~\ref{sec_sol_nr}). In this case, final morphology depends neither on plating current nor on SEI or electrolyte resistivity, but is solely determined by initial nucleation density and the total plated lithium volume. We estimate that this no-ripening regime occurs when the total charge-transfer resistance satisfies $(\Re + \omega)/\kappa \gg \frac{RT}{iF} \approx 26\ \Omega\cdot\text{cm}^2$ for $i = 1$ mA/cm$^2$ at 300 K, by comparing the surface energy term with the charge-transfer resistance and assuming a characteristic nucleus size at nucleation of $\rho^* \sim 1$ (or $r^* \sim 18$ nm for lithium). Given that SEI and electrolyte resistance is often relatively low, this suggests that ripening is likely active under most practical conditions and becomes even more prominent at elevated temperatures, which indirectly was observed in Ref~\cite{Wang2019}.

\begin{table}[!h]
    \centering
    \caption{ Asymptotic behavior of nucleation density $N(t)$, characteristic nucleus radius $r(t)$, and distribution function $f(r,t)$ under different growth regimes $\left(c={\frac{V_m}{vNF}}\right)$. For detailed derivations, see Sec.~\ref{sec_sol}. }
    \begin{tabular}{c c c c}
    \hline\hline
    Conditions&$N(t)$&$r(t)$&$f(r,t)$
    \\\hline
         \multicolumn{4}{c}{No ripening:}\\ 
         \quad$(\Re +\omega)/\kappa\gg\frac{RT}{iF}$\quad&\quad$N\approx const$ \quad&\quad$r\sim  \sqrt[3]{it}$\quad &\qquad \qquad$f\approx f_0\left(r_0\right) \left(\frac{r}{r_0}\right)^{n-2},\quad r_0^{n-1}=r^{n-1}-(cit)^\frac{n-1}{3}$ \\\hline
         
         \multicolumn{4}{c}{3D ripening:}\\ 
         \quad$\Re\ll\omega \rho$\quad&\quad$N\approx const$ \quad&\quad$r\sim  \sqrt[3]{it}$\quad &\quad$f^\infty_{3D}=N\delta(r-\sqrt[3]{cit})$\quad\\\hline
         
         \multicolumn{4}{c}{2D ripening:}\\ 
         \quad$\Re\gg\omega \rho$\quad&\quad$N\sim {i}/{\sqrt{t}}$ \quad&\quad$r \sim \sqrt{t}$\quad &\quad$f^{\infty}_{2D}=5.19 j \frac{\rho}{ (\sqrt{2\tau}-\rho)^3} \exp{\left(\frac{\sqrt{2\tau}}{\rho-\sqrt{2\tau}}\right)}$\quad\\\hline
         \hline
    \end{tabular}
    
    \label{tab_1}
\end{table}

For the converse limiting case of low or moderate total charge-transfer resistance, the system’s behavior critically depends on the dimensionality $n$. In the second limiting case, the 3D ripening regime (low SEI resistance, $\Re\ll \omega \rho$), the size distribution narrows over time and ultimately collapses into a delta function (see $f^\infty_{3D}$ in Tab.\ref{tab_1}, and Sec.~\ref{sec_sol_3D}). Simultaneously, the nuclei density $N$ remains constant while the mean nucleus volume grows linearly with time, $\langle \rho^3 \rangle \sim j\tau$\cite{Scharifker1983}.
This behavior has been observed in galvanic metal deposition without SEI~\cite{Schneider2017}, and is often described as narrowing of the size distribution at later stages of metal deposition\cite{Fransaer1999}.

Lastly, for the 2D regime (high SEI resistivity, $\Re\gg \omega \rho$), which is, as will be motivated later, most relevant for lithium batteries, the distribution function takes a self-similar form and becomes independent of the initial distribution (see Tab.\ref{tab_1}, and Sec.~\ref{sec_sol_2D}). The nuclei density similarly does not depend on the initial nucleation density and follows the relation: 
\begin{equation}
\label{eq_n_2D}
    N\approx1.35 \frac{j}{\sqrt{\tau}}(\rho/r)^2\approx 2.48 \frac{F^2 }{V_m^2}\left(\frac{ R_{SEI} S_{tot}}{ \sigma  }\frac{1}{s} \right)^\frac{3}{2} \frac{i\sqrt{v}}{\sqrt{t}},
\end{equation}
while the mean nucleus size is given by
\begin{equation}
\label{eq_r_2D}
    \langle r\rangle\approx0.844\sqrt{\tau} (r/\rho)\approx 0.689  \frac{V_m}{F} \sqrt{\frac{t\sigma} {R_{SEI}S_{tot}} \frac{s}{v} } .
\end{equation}

Importantly, the expected nucleus radius is independent of the current density but is determined by the metal surface energy and $R_{SEI}$. The nucleation density depends on all three parameters and decreases with time, which at the same current ensures faster growth of the nucleus size $r$  compared to the 3D case (Fig.\ref{fig_2}(b)). Additionally, both nucleation density and growth rate depend on the metal/electrode contact angle $\theta$, which determines the nucleus surface-to-volume ratio. The growth rate reaches its minimal value at $\theta = 120^\circ$ (Fig.\ref{fig_2}(c), as the smaller contact angles ensure denser lithium plating and larger nucleus sizes~\cite{Ye2022}. We remark, that in order to compare with experimental data, the apparent nucleus size should be rescaled as $r_{\text{apparent}} = \alpha r $, where $\alpha=\sin\theta$ for contact angles $\theta < 90^\circ$ and $\alpha=1$ for $\theta \geq 90^\circ$ (Fig. \ref{fig_1}) .

As noted above, the system is expected to evolve from a 2D to a 3D regime, during which the nucleus size distribution undergoes a distinct transformation. Initially, in the 2D regime, the distribution broadens, leading to a wide range of nucleus sizes (left part of Fig.\ref{fig_2}(d)). However, as the  bulk electrolyte resistivity becomes more dominant compared to the SEI resitivity, the distribution gradually narrows and in the asymptotic limit, the distribution approaches a delta function, characteristic of classical diffusion-controlled deposition (right part of Fig.\ref{fig_2}(d)). 

The characteristic moment of regime switching can be estimated from the condition $\rho_{sw} \omega = \Re$. In lithium batteries it is estimated to range from $100$ $\mu$m to $10$ mm, based on values of $R_{SEI}$ and $D$ from Refs.~\cite{Solchenbach2021, PARK201870, Pei2017}. However, this range appears unreasonably large compared to the characteristic feature sizes typically observed in electrode morphology studies (which are usually on the order of $\sim10^0-10^1$ $\mu$m). As a result, lithium batteries are expected to operate purely in the 2D regime, and growth under ripening proceeds until the electrode surface is nearly fully covered and nuclei can no longer be treated as isolated, but rather as a fused, porous medium.

The nucleus size at this full-coverage point ($r_{cov}$) determines the ``weakest point'' in the electrode morphology in terms of coverage density, which directly affects battery longevity. In the SEI-passivated column model~\cite{D0TA06020H}, a simplified approach suggests that the quantity of parasitic side products is approximately proportional to the total exposed metal surface area. As a result, the Coulombic inefficiency, $1 - \text{CE}$, can be estimated as being proportional to the nuclei surface-to-volume ratio, i.e., $\sim 1/r_{cov}$. The upper limit of $r_{cov}$ can be estimated from dense packing as $\pi\alpha^2\int_0^\infty f \rho_{cov}^2 d\rho\approx \frac{\pi}{2\sqrt{3}}$, which in 2D regime corresponds to the time moment $\alpha^2\sqrt{\tau_{cov}}\approx 0.278 j^{-1}$. This, in turn, allows us to estimate the characteristic nucleus radius~$r_{cov}$:
\begin{equation}
	\label{eq_r_cov}
    \begin{aligned}
            t_{cov}\approx& \ 0.0773\frac{1}{\alpha^4 j^2}(t/\tau)\approx 0.0515 \frac{\sigma }{ i^2R_{SEI} S_{tot}}\frac{vs}{\alpha^4},\\
		\langle r_{cov}\rangle\approx&\ 0.844 \sqrt{\tau_{cov}} (r/\rho) \approx 0.156 \frac{\sigma V_m s}{F iR_{SEI} S_{tot} \alpha^2}.
            \end{aligned}
\end{equation}

We deduce that lower current, SEI resistivity, and contact angle all contribute to a larger $\langle r_{cov} \rangle$, resulting in lower surface area exposure, higher CE, and better cycling stability — consistent with experimental observations~\cite{Ye2022,Wang2019}. While nucleation overpotential (NOP) is often highlighted as a key determinant of battery longevity, it does not explicitly appear in our framework. Nevertheless, we imply that NOP is indirectly governed by nucleus geometry: lower $\theta$ results in wider and flatter nuclei and higher electrode surface coverage for the same deposited lithium volume, which reduces local current densities at nuclei surfaces $i_n$. Lower $i_n$ subsequently decreases the overpotential, promoting formation of a more conductive SEI and reducing the final SEI resistivity ($R_{SEI}$)~\cite{Zhang2024,Liu2014}.

\subsection*{Model's limitations}
Our model is constructed under idealized conditions: a flat, homogeneous electrode surface, a uniform SEI layer, and relatively low current densities that preclude dendrite formation. These assumptions define the principal limitations of our theoretical framework.

The first key limitation is the requirement of relatively low current density to avoid tip-enhanced growth, which can ultimately lead to dendrite formation. Specifically, the model assumes that ion transport is sufficiently fast to maintain quasi-equilibrium over the characteristic size of a growing nucleus. This condition holds when the local current density at the electrode surface remains significantly below the diffusion-limited current density~\cite{Barton1962,Despic_1968}.

The second limitation arises from SEI inhomogeneity. In practice, the solid electrolyte interphase may exhibit spatial variations in thickness, composition, and ionic conductivity, stemming from initial electrode surface irregularities or from mechanical degradation such as cracking. These variations can lead to uneven local current distribution and non-uniform lithium deposition. Notably, lower initial SEI resistivity has been shown to reduce the likelihood of subsequent SEI disruption~\cite{KO2023232779}. However, it is important to emphasize that if SEI inhomogeneities occur on a length scale smaller than the typical nucleus size $\rho_s$ (and do not induce dendritic growth), our model remains applicable, subject to corrections to the geometric shape factors $s$ and $v$. 

The third limitation involves substrate inhomogeneity. Even under a spatially uniform electric potential across a metallic substrate, local variations in crystallographic orientation, defects, or grain boundaries can influence the initial SEI formation and lithium nucleation. While these effects can be pronounced during the nucleation stage, their impact on subsequent growth diminishes if the contact angle $\theta$ remains uniform across the surface, since the growth phase is determined by nucleus, SEI, and electrolyte properties. In cases of mild initial inhomogeneity, the initial perturbations typically average out during growth, allowing the model to remain applicable.
In contrast, in strongly inhomogeneous systems—such as passivated current collectors where lithium nucleation is confined to rare, defective regions of a passivation layer~\cite{Du2024}—the initial nucleus density can be extremely low for ripening process, and system remains in non-ripenning regime throughout deposition.

Finally, we note that lithium redistribution could, in principle, occur along the electrode surface; however, this pathway is unlikely to be significant beyond the initial nucleation stage. Although a few reports have considered such on-surface migration (e.g.,~\cite{Wu2023}), its efficiency is expected to be low: for surface migration, the effective cross-sectional area of the transport channel scales as $2\pi r h$, where $h$ is on the order of the Van-der-Waals diameter of a lithium atom ($\approx 2.24$ \AA), whereas the surface area of a nucleus scales as $\sim \pi r^2$. Unless the lithium diffusion coefficient along the electrode surface is several orders of magnitude greater than through the SEI—which seems improbable given that the surface is typically fully passivated before nucleation~\cite{Xu2023}—this redistribution channel is unlikely to play a dominant role beyond the earliest stages. 

\subsection*{Comparison with experiments}

Our predictions qualitatively align with previous experimental observations, particularly regarding the effects of SEI resistivity on nucleus size, $\langle r_{cov}\rangle$, and consequently battery longevity. In addition, the observed decrease in nucleation density~\cite{Pei2017} is predicted by our approach, but cannot be explained by the standard nucleation-and-growth formalism~\cite{Scharifker1983}. To further validate our model quantitatively , we directly compared the predicted average nucleus size, size distribution and coloumbic efficiency to experimental data. Unfortunately, there are not many studies that investigate all aspects required for comparison with the models result. We identified two studies~\cite{Wang2019,Pei2017} that provide extensive experimental data, making quantitative validation possible. Both studies used a 1M LiTFSI in DOL/DME electrolyte, with one varying cell temperature~\cite{Pei2017} and the other varying current density and deposited charges~\cite{Wang2019}.

One of the most challenging aspects of experimental interpretation is determining the correct value of $R_{SEI}$. Standard electrochemical impedance spectroscopy (EIS) cannot distinguish $R_{SEI}$ from other resistive elements in the equivalent circuit~\cite{Solchenbach2021}, leading to significant discrepancies. The true value of $R_{SEI}$ can be several orders of magnitude lower than the total charge transfer resistance measured by EIS. For example, in a 1M LiPF$_6$ in EC/EMC electrolyte, $R_{SEI}$ measured under blocking conditions is 1.5 $\Omega\cdot$cm$^2$\cite{Solchenbach2021}, whereas conventional EIS reports a total resistance of approximately 100~$\Omega\cdot$cm$^2$\cite{Hobold2023}. To minimize this inconsistency, we assumed that $R_{SEI}$ follows an Arrhenius relation, $R_{SEI} = A\exp\left(\frac{E_A}{RT}\right)$, where $E_A$ was fitted to experimental data ($E_A = 32$ kJ/mol, see Fig.\ref{fig:R_ct_fit}), leaving $A$ as the only free parameter in our model. This approach yielded consistent and reproducible results, independently converging to $R_{SEI} \approx 0.27$ $\Omega\cdot$cm$^2$ at 300~K\footnote{\label{fn1}This value likely underestimates $R_{SEI}$ due to the omission of small nuclei in the original size histograms. While this discrepancy has a negligible effect on the $\langle r \rangle$, it can lead to an overestimation of $\langle r^3 \rangle$, and thus an underestimation of $N$ by up to a factor of five.} for both referenced studies\cite{Wang2019,Pei2017}.

 \begin{figure}[!h]
     \centering
     \includegraphics[width=0.75\linewidth]{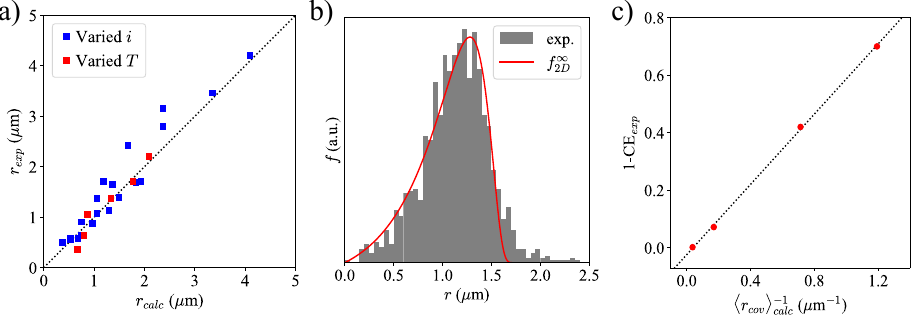}
     \caption{\textbf{Comparison of model predictions (Eqs.\ref{eq_n_2D}-\ref{eq_r_cov}) with experimental data.} (a) Nucleus radii ($r$) at different temperatures ($T$)~\cite{Wang2019} are shown as red squares. Temperature-dependent variations in $R_{\mathrm{SEI}}$ were obtained via Arrhenius fit of measured values (see Section~\ref{sec_exp_fit}). Blue squares represent nucleus radii at different current densities ($i$) and deposited charges~\cite{Pei2017}. (b) Experimental nucleus size distribution extracted from a large-area SEM image (Fig.S6(a)\cite{Pei2017}) is shown as a bar plot; the red line depicts the theoretical 2D distribution $f^\infty_{2D}$ (Tab.\ref{tab_1}).
     (c) Experimentally measured coulombic inefficiency (1-CE)~\cite{Wang2019} vs. calculated $1/r_{cov}$ (Eq.~\ref{eq_r_cov}). 
     }
     \label{fig_3}
 \end{figure}

Predicted and experimentally observed average nuclei sizes are in excellent agreement over a wide temperature range (–20 $^\circ$C to 30 $^\circ$C) and plating currents densities (0.1–5 mA/cm$^2$) (Fig.~\ref{fig_3}(a)). The model also reproduces the size distribution, capturing the characteristic tail in the small-nucleus region (Fig.~\ref{fig_3}(b)), while the slight mismatch in the large-nucleus tail is likely attributable to coalescence effects~\cite{PhysRevB.81.075319}. Notably, we emphasize the clear linear correlation between Coulombic inefficiency (1–CE)  and the inverse of the calculated characteristic radius at the point of full electrode coverage, $1/r_{cov}$ (Fig.~\ref{fig_3}(c), see the discussion above Eq.\ref{eq_r_cov}). Overall, the strong correlation between theoretical and experimental results validates the accuracy and predictive power of our model.

\section*{Summary \& Conclusion}
In this work, we developed a theoretical framework to describe the evolution of electrode morphology under the competing influences of electroplating and surface energy–driven redistribution. The analysis reveals three key asymptotic regimes. At high currents and large charge-transfer resistance, ripening is negligible and nuclei volumes grow uniformly without material redistribution. In contrast, at low charge-transfer resistance and current density, Ostwald ripening dominates: in the surface-resistance-limited 2D regime this leads to broader size distributions and decreasing nuclei densities, whereas in the bulk-diffusion-controlled 3D regime the distribution gradually narrows under constant nucleus density. While lithium batteries are expected to operate mainly in the 2D regime, galvanic systems with lower (or without) SEI resistivity are instead well described by the 3D asymptote and may even exhibit the distinct transition from 2D to 3D behavior.

We highlight in particular the analytical expressions derived for the 2D regime. These relations connect nucleus size, size distribution, and density to experimentally accessible parameters such as $R_{\text{SEI}}$, plating current, and electrode wettability. Notably, the framework predicts the characteristic nucleus radius ($r_{cov}$) and the coverage time ($t_{cov}$) at which the electrode surface approaches full coverage. The latter correlates directly with Coulombic efficiency (CE), following $1 - \text{CE} \sim 1/r_{cov}$. Validation of our theoretical predictions against two independent experimental datasets showed excellent agreement. Taken together, our results establish a quantitative link between key experimental variables and lithium plating morphology, CE, and ultimately battery lifetime. In practical terms, this translates into the following relationships:


\begin{itemize}
    \item \textbf{Electrode wettability}, determined by the metal/electrode contact angle $\theta$, controls the nuclei surface-to-volume ratio. Lower $\theta$ increases surface coverage for the same deposited charge, reduces effective current density at the nuclei surface, and in turn lowers nucleation overpotential and SEI resistivity. While the dependence becomes non-monotonic for $\theta \geq 90^\circ$, variations in this range have a minor impact compared with the strong improvement observed for $\theta < 90^\circ$, which approaches ideal surface coverage as $\theta \to 0$.

    \item \textbf{SEI resistivity ($R_{SEI}$) and current density ($i$)} jointly determine nuclei sizes and density. Varying both affects the electrode coverage time as $t_{cov} \sim (i^2 R_{SEI})^{-1}$. The morphology at this point has a strong impact on Coulombic efficiency and lifespan, and is highly susceptible to increases in current density ($1 - \mathrm{CE} \sim i R_{SEI}$). After coverage, sensitivity to $i$ decreases, allowing higher currents with less impact on performance. These results provide an analytical framework for optimizing charging protocols and estimating the trade-off between Coulombic efficiency and charging speed.
\end{itemize}

Although our study focused on lithium metal, the underlying principles apply broadly to metal electrodeposition systems where surface morphology plays a critical role. We anticipate that the analytical model presented here will support more precise optimization of battery parameters, ultimately contributing to improved battery performance and lifetime.

\printbibliography
\clearpage

\section*{Table of Contents}
\begin{center}
  \includegraphics[width=5cm]{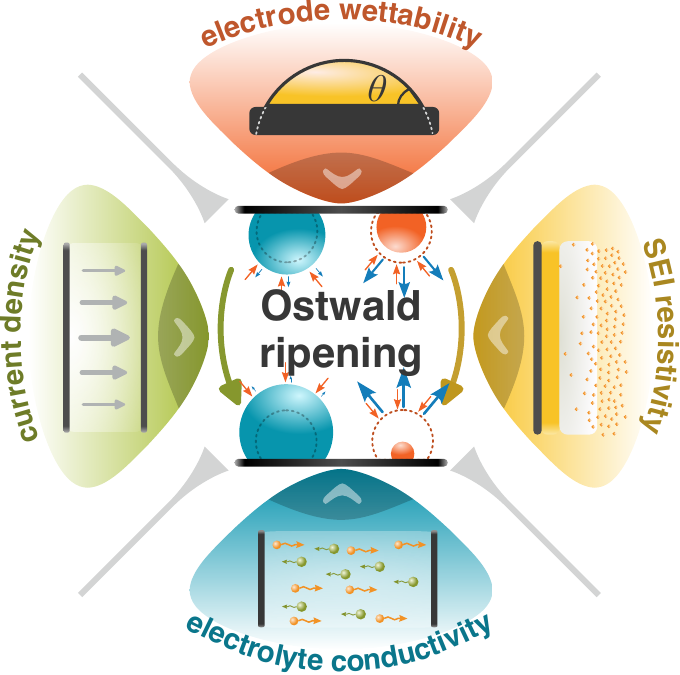}  
\end{center}

This study introduces a predictive theory for electrochemical Ostwald ripening during lithium metal electrodeposition. The model identifies surface energy–driven redistribution as a key factor shaping lithium morphology and determining Coulombic efficiency in flat-anode cells. Analytical predictions match experimental trends and offer practical guidelines for improving plating uniformity and battery performance across electrochemical systems.

\clearpage
\renewcommand{\thesection}{S\arabic{section}}
\renewcommand{\thetable}{S\arabic{table}}
\renewcommand{\thefigure}{S\arabic{figure}}
\renewcommand{\theequation}{S\arabic{equation}}
\renewcommand{\thepage}{S\arabic{page}}
\setcounter{figure}{0}
\setcounter{section}{0}
\setcounter{table}{0}
\setcounter{equation}{0}
\setcounter{page}{1}
\begin{center}
\section*{Supplementary Information: \art_title}

\authors\\
$^\dagger$Institute of Physics, Ecole Polytechnique Fédérale de Lausanne (EPFL), CH-1015 Lausanne, Switzerland
\end{center}

\tableofcontents

\section{Units and quantities}\label{sec_units}

\hspace*{1.5em}$\theta$ - metal/electrode contact angle

$s=2\pi (1-\cos\theta)$ - nucleus surface coefficient. exposed nucleus surface is $s r^2$

$v= \frac{\pi}{3}(2+\cos\theta)(1-\cos\theta)^2$ - nucleus volumetric coefficient. Total nucleus volume $vr^3$

$\sigma$ - metal/electrolyte surface energy. For lithium $\sigma=1.716$ J/m$^2$~\cite{Monroe_2003}

$ V_m$ - metal's molar volume. For lithium $ V_m=13$ cm$^3$/mol~\cite{NIST_WTT}

$\Delta G=G_e + \frac{3\sigma V_m}{r}$ - total nucleus free energy, $G_e$ - environmental Gibbs energy

$F=96.485$ s A/mol - Faraday constant

$\eta$ - overpotential

$i$ - cell current density

$i_n$ - current density at nucleus surface

$i_0$ - electrode exchange current

CE - Coulombic efficiency

$R_{SEI}$ - SEI resistance

$S_{tot}$ - total electrode surface

$R_{SEI}'=R_{SEI}+\frac{RT}{i_0F S_{tot}}\approx R_{SEI}$ corrected SEI resistance

$D$ diffusion coefficient of metal ions in electrolyte

$C_\infty$ concentration of metal ions in bulk electrolyte

$\kappa$ surface conductance $\kappa^{-1}\approx R_{SEI}' S_{tot}$ in the SEI diffusion-limited regime, and $\kappa^{-1}\approx \frac{2\sigma V_m}{D C F^2 }$ in the electrolyte 
\hspace*{1.1em} diffusion-limited regime.

$\Re=\kappa R_{SEI}' S_{tot}$ - dimensionless SEI resistance

$\omega=\frac{2\sigma V_m\kappa}{D C_\infty F^2 }$ - dimensionless electrolyte resistivity

$N$ - nuclei density

$r$ - nucleus radius

$r^*\sim\frac{2\sigma V_m}{RT}$ - critical radius

$r_{apparent}=\alpha r$ - apparent nucleus radius, where $\alpha=1$ for $\theta\geq90^\circ$, and $\alpha=\sin\theta$ for $\theta<90^\circ$

$\rho= \frac{RT}{2\sigma V_m}r$  - dimensionless radius

$r_s$  - minimal stable nuclei radius

$\rho_s=\frac{RT}{F\eta}$  - dimensionless minimal stable nuclei radius

$r_{cov}$ - the nucleus radius at which the electrode surface becomes completely covered

$t_{cov}$ - the corresponding timestep 

$f$ - nuclei size distribution function

$v_\rho=\frac{d \rho}{d\tau}$ - nucleus growth rate

$\tau=\frac{\kappa(RT)^2 }{6 \sigma F^2 } \frac{s}{v} t$ dimensionless time

$j =  \frac{F i}{\kappa RT}\frac{3}{s}$ dimensionless flow

$\nu = N \left(\frac{2\sigma V_m}{RT}\right)^2$ - dimensionless nuclei density

\section{Solution for electrochemical Ostwald ripening}
\label{sec_sol}
As outlined in the main text, the evolution of the nucleus size distribution during electrochemical Ostwald ripening is governed by the following system of equations.\\
The equation of motion
\begin{equation}
\label{seq_d_rho_d_tau_n}
v_\rho=\frac{d\rho}{d\tau}=\frac{1}{\rho^{{n-1}}}\left(\frac{\rho}{\rho_s}-1\right),
\end{equation}

determines how nuclei with radius $\rho$ evolve in time. Nuclei smaller than the characteristic radius $\rho_s$ shrink, while nuclei bigger than the characteristic radius will grow. The evolution of the characteristic radius $\rho_s$ with time is determined in accordance with the mass conservation Eq.\ref{seq_consv_of_mat_general}. \\
The continuity equation
\begin{equation}
\label{seq_f_cont}
\frac{d f}{d\tau}+\frac{d}{d \rho}\left(f v_\rho\right)=0,
\end{equation}

is mathematically equivalent to the equation of motion. The time evolution of individual particles with radius $\rho$ is rewritten as the time evolution of the radius distribution function $f$.\\
The mass conservation condition under a constant current
\begin{equation}
\label{seq_consv_of_mat_general}
 j\tau=\int_0^\infty f(\rho,\tau) \rho^3d\rho,\qquad \text{or} \qquad j=3\int_0^\infty f(\rho,\tau)v_\rho \rho^2d\rho. 
\end{equation}
corresponds to metal plating when $j \geq 0$. Note that, equations \ref{seq_d_rho_d_tau_n} and \ref{seq_f_cont} are the same as in Lifshitz-Slyozov~\cite{Lifshitz1961} and Wagner~\cite{Wagner1961}, while the constant current assumption \ref{seq_consv_of_mat_general} consitutes the original part of this work. Differences to the classic constant volume assumption are derived and commented on in \ref{sec_sol_2D} and \ref{sec_sol_3D}. \\

\subsection{No ripening}
\label{sec_sol_nr}

In the absence of ripening, we assume $\rho_s \ll \rho$, which simplifies the equation of motion (Eq.~\ref{seq_d_rho_d_tau_n}) and yields:
\begin{equation}
\rho^{n-1} = \rho_0^{n-1} + (n-1)\int_0^\tau \frac{d\tau'}{\rho_s(\tau')} = \rho_0^{n-1} + h^{n-1}(\tau),
\label{seq_rho_t_nr}
\end{equation}
where $\rho_0$ is the initial nucleus radius corresponding to a nucleus of size $\rho$ at time $\tau$.

In this regime, the size distribution shifts toward larger $\rho$ values over time, while conserving the nucleus number density:
\[
\nu = \int f(\rho,\tau) \, d\rho = \int f_0(\rho_0) \, d\rho_0 = \nu_0 = \text{const},
\]

From Eq~\ref{seq_rho_t_nr}, we have $d\rho /d \rho_0 = \left(\frac{\rho_0}{\rho}\right)^{n-2}$, which using $f(\rho, \tau) d\rho = f_0(\rho_0) d\rho_0$ leads to 
\begin{equation}
\label{seq_f_nr}
f(\rho, \tau) = f_0(\sqrt[n-1]{\rho^{n-1}-h^{n-1}(\tau)}) \cdot \left(\frac{\rho}{\sqrt[n-1]{\rho^{n-1}-h^{n-1}(\tau)}}\right)^{n-2}.
\end{equation}
In 2D, the size distribution remains unchanged in shape and is simply shifted toward larger radii. In contrast, in the 3D case, the distribution becomes narrower, and the function $f$ assumes bigger values to conserve the total number of nuclei, resulting in a sharper distribution over time.

Mass conservation requires that the total deposited volume equals the integrated current: 
\[
j \tau = \int_0^\infty f_0(\rho_0) \left( \rho_0^{n-1} + h(\tau)^{n-1} \right)^\frac{3}{n-1} d\rho_0.
\]

In the large-time limit, neglecting $\rho_0$ gives the universal scaling law:
\begin{equation}
\label{seq_rho_nr}
\rho\approx h(\tau) \approx \left( \frac{j\tau}{\nu_0} \right)^{1/3},
\end{equation}

\subsection{The 2D ripening case} 
\label{sec_sol_2D}
Analogously to the Lifshitz–Slyozov approach, we introduce the variable transformation $z = \rho / \rho_s$ and $\xi = \ln{\rho_s^2}$. Here, $z$ represents the rescaled nucleus radius, normalized by the evolving characteristic radius $\rho_s$, while $\xi$ taken as the logarithm of the growing characteristic radius $\rho_s$ acts as the new time variable. Since $\rho_s \to \infty$ as $\tau \to \infty$, the asymptote $\xi \to \infty$ exists and the asymptotic behavior ($\tau \to \infty$ and equivalently $\xi \to \infty$) of the system will be the focus of the following discussion. \\

As a first step, we rewrite the equation of motion (Eq.~\ref{seq_d_rho_d_tau_n}) for the 2D case in terms of the rescaled variable $z$

\begin{equation}
    \label{seq_of_motion_z_2D}
    v(z, \gamma) = \frac{\partial z}{\partial \xi} = \gamma\left(1- \frac{1}{z}\right) -\frac{1}{2}z,
\end{equation}

where $\gamma = \frac{d\tau}{d \rho_s^2}$. Denoting $\phi(z,\xi)$ as the new distribution function and preserving the distribution density $(\phi(z,\xi)dz = f(\rho,\tau)d\rho)$, the continuity equation (Eq.~\ref{seq_f_cont}) transforms into

\begin{equation}
    \label{seq_continuty_equation_z}
    \frac{\partial \phi}{\partial \xi} + \frac{\partial}{\partial z} (\phi v_z) = 0.
\end{equation}

In the new coordinates ($z$ and $\xi$) the mass conservation Eq.\ref{seq_consv_of_mat_general} can be rewritten as 

\begin{equation}
\label{seq_consv_of_mat_z_2D}
    j = 3e^{\frac{\xi}{2}} \int_0^\infty (z^2 -z) \phi dz
\end{equation}

Interpreting $\gamma$ as the new $\rho_s$, the analogy between Eq. \ref{seq_of_motion_z_2D}, \ref{seq_continuty_equation_z}, \ref{seq_consv_of_mat_z_2D}
and Eq. \ref{seq_d_rho_d_tau_n}, \ref{seq_f_cont}, \ref{seq_consv_of_mat_general}, $z$ and $\rho$ as well as $\xi$ and $\tau$ is completed. Indeed $\rho_s$ and $\gamma$ both serve as global parameters in the equation of motion  and their time evolution $\gamma(\xi)$ and $\rho_s(\tau)$ is dictated by the mass conservation condition.  The asymptotic limit of $\gamma(\xi)$ as $\xi \to \infty$ is however much 
more convenient to analyze compared to the limit $\rho_s(\tau)$ as $\tau \to \infty$, as we will see next. \\

\begin{figure}[h!]
    \centering
    \includegraphics[width=0.5\linewidth]{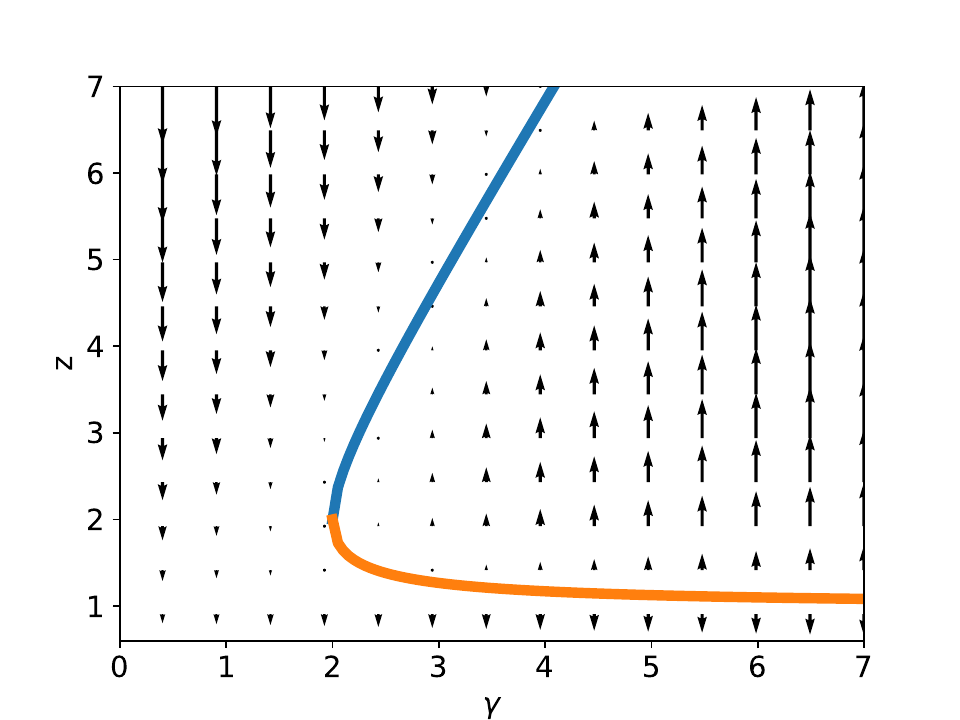}
    \caption{Flow diagram of the rescaled nucleus size $z$ as governed by Eq.~\ref{seq_of_motion_z_2D}. Arrows indicate the direction of $v(z, \gamma)$ for different values of $\gamma$. The orange line denotes the unstable (repelling) fixed point $z_-$, while the blue line marks the stable (attracting) fixed point $z_+$.}
    \label{sfig:flow_diag_2D}
\end{figure}

 We begin our discussion with the equation of motion Eq.\ref{seq_of_motion_z_2D}, which governs how individual nuclei evolve over time. The corresponding flow diagram, with $\gamma$ treated as a free parameter, is shown in Fig.\ref{sfig:flow_diag_2D}. We observe that the diagram can be divided into two different regimes. For $\gamma < 2$, all nuclei regardless of their size $z$ decrease with time $\xi$, while for the $\gamma > 2$ regime, nuclei sizes greater than $z_- = \gamma - \sqrt{\gamma^2- 2\gamma}$ (orange line in Fig~\ref{sfig:flow_diag_2D}) grow and eventually converge to $z_+ = \gamma + \sqrt{\gamma^2- 2\gamma}$ (blue line in Fig~\ref{sfig:flow_diag_2D}), while nuclei sizes below the orange line shrink and vanish. \\

 To further fulfill the mass conservation $\gamma(\xi)$ will evolve as a function of $\xi$. We anticipate the result from the following slightly technical proof and state that when $\gamma < 2$ the dissolution of nuclei is so severe that a constant current plating cannot be sustained. This prompts $\gamma$ to increase to values $\geq 2$. On the contrary if $\gamma > 2$, nuclei will grow too fast leading to a faster than linear current, such the evolution of $\gamma$ sees it oscillating around the value of $2$ with $2$ being the only asymptotic value possible. \\

\begin{figure}[h!]
    \centering
    \includegraphics[width=0.5\linewidth]{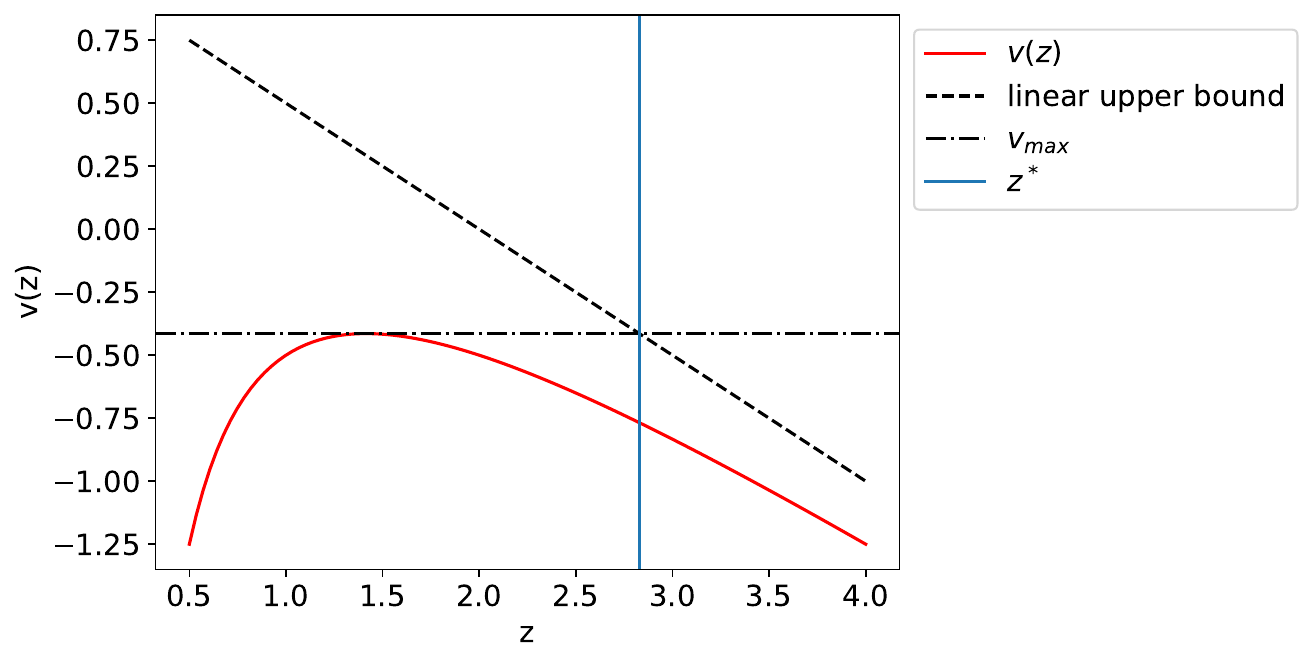}
    \caption{Upper bounds of the function $v(z)$ for the case $\gamma < 2$. 
    For $z < z^*$, $v(z) \leq v_{\mathrm{max}}$, whereas for $z > z^*$, $v(z) \leq \gamma - \tfrac{1}{2}z$.}
    \label{sfig:vz_2D}
\end{figure}

We start the technical proof by first taking a closer look at the $\gamma < 2$ case. We define the maximum value of $v(z)$ as $v_{max}=\gamma-\sqrt{2\gamma}$, and $v_{max}<0$. Furthermore $v(z)$ can be bounded from above by a linear function $v(z) \leq \gamma-\frac{1}{2}z$ (see Fig \ref{sfig:vz_2D}). As a consequence we write

\[ \frac{\partial z}{\partial \xi} = v(z) \leq \gamma - \frac{1}{2}z, v_{max} \]

and nuclei dissolve at least as fast as both, the exponential decay $z(\xi) \leq 2\gamma + (z_0-2 \gamma)e^{-\frac{1}{2}\xi}$ and the linear decrease $z(\xi) \leq z_0 - v_{max} \xi$, where $z_0$ is the initial nucleus size at $\xi = 0$. Combining both gives a stronger upper bound 

\[ \frac{\partial z}{\partial \xi} = v(z)  \leq \min\{\gamma - \frac{1}{2}z, v_{max}\} < 0\]

which sees nuclei bigger than $z^*$ dissolving at least exponentially and nuclei smaller than $z^*$ at least linearly (see Fig\ref{sfig:vz_2D}). The nucleus size $z^*$ is chosen as the intersection of $v_{max}$ and $\gamma-\frac{1}{2}z$ and is given by $z^* = 2\sqrt{2 \gamma} > 2 \gamma$.  \\

We now use this upper bound on the nucleus size, to show that keeping $\gamma < 2$ effectively leads to a deplating of metal contradicting in particular the constant plating current assumption with $j \geq 0$. Let $\phi_0(z_0)$ denote the initial distribution at $\xi = 0$ and let $z(z_0, \xi)$ denote the evolution of an initial nucleus of size $z_0$ with time. Then 
\begin{align*}
    j =& 3e^{\frac{\xi}{2}} \int_0^\infty \phi (z^2 -z)  dz = 3e^{\frac{\xi}{2}} \int_0^\infty \phi z^2dz -3e^{\frac{\xi}{2}} \int_0^\infty \phi zdz
\end{align*}

Since the second integral $I_2$ is $\geq 0$, it is sufficient to show that the first integral $I_1$ vanishes in the large-time limit. We write
\begin{align*} 
I_1 &=3e^{\frac{\xi}{2}} \int_0^\infty \phi z^2  dz = 3 e^{\frac{\xi}{2}} \int_0^\infty \phi_0(z_0) z(z_0, \xi)^2dz_0 < 3 e^{\frac{\xi}{2}} \int_{(z^*-2\gamma)e^{\frac{1}{2} (\xi-\xi_0)}}^\infty \phi_0(z_0) (2 \gamma + (z_0 - 2\gamma) e^{-\frac{1}{2} \xi})^2 dz_0.
\end{align*}

The inequality uses a direct application of the exponential decay upper bound $v(z) < \gamma -\frac{1}{2}z$ for the integrand, while  the lower integration bound additionally subtracts nuclei with initial nucleus size $z_0$, which after a time $\xi$ are completely dissolved. To estimate the lower bound for $z_0$, assume that $\xi_0$ denotes the time to linearly dissolve a nucleus of size $z^*$, then the remaining time $\xi -\xi_0$ can be used to dissolve a nucleus of size of at least $z_0 = 2\gamma + (z^*-2\gamma)e^{\frac{1}{2} (\xi-\xi_0)}$ to $z^*$. For simplicity the slightly smaller lower bound $z_0 = (z^*-2\gamma)e^{\frac{1}{2} (\xi-\xi_0)}$ is then taken.\\

Expanding 

\begin{align*}
    &3 e^{\frac{\xi}{2}} \int_{(z^*-2\gamma)e^{\frac{1}{2} (\xi-\xi_0)}}^\infty \phi_0(z_0) (2 \gamma + (z_0 - 2\gamma) e^{-\frac{1}{2} \xi})^2 dz_0 \\
    &=  3e^{\frac{1}{2} \xi} \int_{(z^*-2\gamma)e^{\frac{1}{2} (\xi-\xi_0)}}^\infty 2  \gamma^2  \phi_0(z_0)dz_0  \\
    &+ 3\int_{(z^*-2\gamma)e^{\frac{1}{2} (\xi-\xi_0)}}^\infty  4\gamma(z_0 - 2\gamma) \phi_0(z_0)dz_0 \\
   & + 3e^{-\frac{1}{2}\xi} \int_{(z^*-2\gamma)e^{\frac{1}{2} (\xi-\xi_0)}}^\infty 2  (z_0 - 2\gamma)^2\phi_0(z_0)dz_0
\end{align*} 

the term with prefactor  $e^{-\frac{1}{2}\xi}$ will go to zero when $\xi \to \infty$. Similarly the term with constant prefactor  goes to zero as the lower integration  bound approaches infinity (as $z_0 \phi_0(z_0)$ is an integrable function\footnote{\label{sfn1}The functions $\phi_0(z_0) z_0^n$ for $n\in[0,3]$ is integrable, as these terms correspond to physical quantities such as nuclei density, current, and volume at $\xi = 0$, which are assumed to be finite.}). Finally the term with prefactor $e^{\frac{1}{2} \xi}$ can be written and upper bounded as 

\[ A x \int_{Bx}^\infty \phi_0(z_0) dz_0 = \frac{A}{B} Bx \int_{Bx}^\infty \phi_0(z_0) z_0 dz_0 < \frac{A}{B} \int_{Bx}^\infty \phi_0(z_0) z_0^2 dz_0 \]

where $A = 6 \gamma^2$, $B = (z^*- 2 \gamma)e^{-\frac{1}{2}\xi_0}$ and $x = e^{\frac{1}{2}\xi}$. Since $\phi_0(z_0) z_0^2$ is an integrable function as well, this term also goes to zero as the lower integration bound goes to infinity. We hence showed that $\gamma < 2$ needs to increase if a constant plating current needs to be maintained. \\

Now we turn our attention to the slightly easier $\gamma > 2$ case.  It turns out that asymptotic solutions exists such that $\gamma$ in principle can stay constant at some value $>2$. The solutions are however unstable, as a perturbation generating a small fraction of nuclei above the orange line $z_- = \gamma - \sqrt{\gamma^2- 2\gamma}$, leads to these nuclei never being dissolved again and growing as $\rho_s \propto \tau^{1/2}$. These fast growing nuclei take up more lithium than can be plated by a constant current ($\tau^{3/2}$ dependence of lithium in these fast growing nuclei versus $\tau$ dependence of lithium plated by constant current) such that to ensure constant current plating $\gamma$ will decrease until it reaches a value $\leq2$. We therefore conclude that due to the mass conservation equation $\gamma$ will evolve towards and asymptotically stabilize at the value $\gamma = 2$, as it is the only dynamically consistent solution. \\

We now derive the time evolution of the nucleus density and the nucleus radius distribution in the asymptotic limit where $\gamma =2$. Note that the $\gamma$ is kept general until equation  Eq. \ref{seq_f_cont_z_timeind} as the same analysis with $\gamma \neq 2$ will be needed for the 3D analysis. \\

Introducing:

\[
    g(z, \gamma) = -v(z, \gamma), \quad\text{and}\quad \psi = \int_0^z g^{-1}(z', \gamma) dz',
\]
the general solution of Eq~\ref{seq_continuty_equation_z} is 

\begin{align*}
    \phi(z, \tau) = \chi(\xi + \psi) \frac{1}{g(z, \gamma)}
\end{align*}

The mass conservation Eq.\ref{seq_consv_of_mat_z_2D} can be rewritten as 

\begin{equation*}
    j = 3e^{\frac{\xi}{2}} \int_0^\infty (z^2 -z) \phi dz = 3e^{\frac{\xi}{2}} \int_0^\infty (z^2 -z) \chi(\xi + \psi) \frac{1}{g(z, \gamma)} dz.
\end{equation*}

Since the left side is time independent we find that 

\[
    \chi(\xi + \psi) = e^{-\frac{1}{2}( \xi + \psi)}
\]

Interestingly, it is at this point that the constant current case deviates from the constant volume 2D Ostwald ripening~\cite{Wagner1961}:

\[
    V = e^{\frac{3\xi}{2}} \int_0^\infty \phi z^3 dz,
\]
which leads to
\[
    \chi(\xi + \psi) = e^{-\frac{3}{2}( \xi + \psi)}.
\]

We generalize both cases by writing 
\begin{align*}
    \phi(z, \xi) = e^{-\alpha \xi} \Phi(z),
\end{align*} 
where $\alpha = \frac{1}{2}$ in the case of constant current and $\alpha = \frac{3}{2}$ in the case of constant volume. Substituting this into the continuity equation (Eq.~\ref{seq_f_cont}) leads to the time-independent form:

\begin{equation}
    \label{seq_f_cont_z_timeind}
    v\frac{\partial \Phi}{\partial z} + \Phi \frac{\partial v}{\partial z} = \alpha \Phi
\end{equation}

Introducing the function $\beta(z) = \alpha - \frac{dv}{dz}$ and setting $\gamma = 2$, we solve the equation using separation of variables:
\[\frac{\beta}{v} dz = \frac{1}{\Phi}{d\Phi},\]
which yields the solution:
\[
\Phi(z) = \frac{z}{(2-z)^{2(\alpha+1)}} e^{\frac{4\alpha}{z-2}}, 
\]
with the physical requirement that $\Phi(z) = 0$ for $z > z_+ = z_- = 2$. Furthermore, it is straightforward to show that the root $z_\beta$ of $\beta(z)$ must satisfy $z_\beta < z_+ = z_- = 2$ for the solution to be physical. This is equivalent to the condition $\alpha > 0$, which holds for both the constant current ($\alpha = \frac{1}{2}$) and constant volume ($\alpha = \frac{3}{2}$) cases.

The final solution for the constant current case is hence 
\begin{align}
\phi(z, \xi) = A \frac{z}{(2 - z)^3} \exp\left(\frac{2}{z - 2}\right) e^{-\xi/2},
\end{align}
which, when transformed back to the original variables $\rho$ and $\tau$, yields:
\[
f^{\infty}_{2D}(\rho, \tau) = A \frac{\rho}{(\sqrt{2\tau} - \rho)^3} \exp\left( \frac{\sqrt{2\tau}}{\rho - \sqrt{2\tau}} \right).
\]

The conservation of mass Eq.~\ref{seq_consv_of_mat_z_2D}, provides a normalization condition for constant current plating

\[
j=3A\int_0^2 (z^2-z) \frac{z}{(2-z)^3} e^{\frac{2}{z-2}} dz \approx 0.1925 A
\]

We derive the following important quantities: 

\begin{itemize}
    \item \textbf{Nucleation density}:
    \[
    \nu = \int_0^2 \phi(z, \xi)\, dz = \int_0^\infty f(\rho, \tau)\, d\rho \approx 0.956\, j\, e^{-\xi/2} \approx 1.35 \frac{j}{\sqrt{\tau}}.
    \]

    \item \textbf{Mean nucleus radius}:
    \[
    \nu \langle \rho \rangle = \int_0^\infty \rho\, f(\rho, \tau)\, d\rho = e^{\xi/2} \int_0^2 z\, \phi(z, \xi)\, dz \approx 1.14\, j \rightarrow \langle \rho \rangle \approx0.844 \sqrt{\tau}.
    \]

    \item \textbf{Mean square radius}:
    \[
    \nu \langle \rho^2 \rangle = \int_0^\infty \rho^2\, f(\rho, \tau)\, d\rho = e^{\xi} \int_0^2 z^2\, \phi(z, \xi)\, dz \approx 1.47\, j\, e^{\xi/2} \approx 1.04\, j\, \sqrt{\tau}  \rightarrow \langle \rho^2 \rangle \approx 0.770 \tau.
    \]
\end{itemize}

We emphasize here that the derivation for the constant current case represents, to the best of our knowledge, a new result. Interestingly, the discussion on the asymptotic value of $\gamma$ is exactly the same as in the constant-volume 2D Ostwald ripening case. This will not remain true for the 3D case. The difference between constant current and constant volume only became apparent in the final calculation of the distribution function $f$, where distinct values $\alpha = \frac{1}{2}$ and $\alpha = \frac{3}{2}$  were used. \\

\subsection{The 3D ripening case} 
\label{sec_sol_3D}

In the 3D case, we again follow the Lifshitz-Slyozov approach~\cite{Lifshitz1961} and define variables $z = \rho/\rho_s$ and $\xi = \ln{\rho_s^3}$.

In a first step we rewrite the equation of motion as 

\begin{equation}
    \label{seq_of_motion_z_3D}
    v(z, \gamma) = \frac{\partial z}{\partial \xi} = \gamma\frac{z-1}{z^2}-\frac{z}{3},
\end{equation}
where $\gamma = \frac{d\tau}{d \rho_s^3}$. The continuity equation retains the same form as in Eq.~\ref{seq_continuty_equation_z}, while the mass conservation condition for constant-current 3D ripening becomes
\begin{equation}
    \label{seq_mass_cons_3D}
    j = 3 \int_0^\infty (z - 1) \phi(z) dz
\end{equation}

\begin{figure}[h!]
    \centering
    \includegraphics[width=0.5\linewidth]{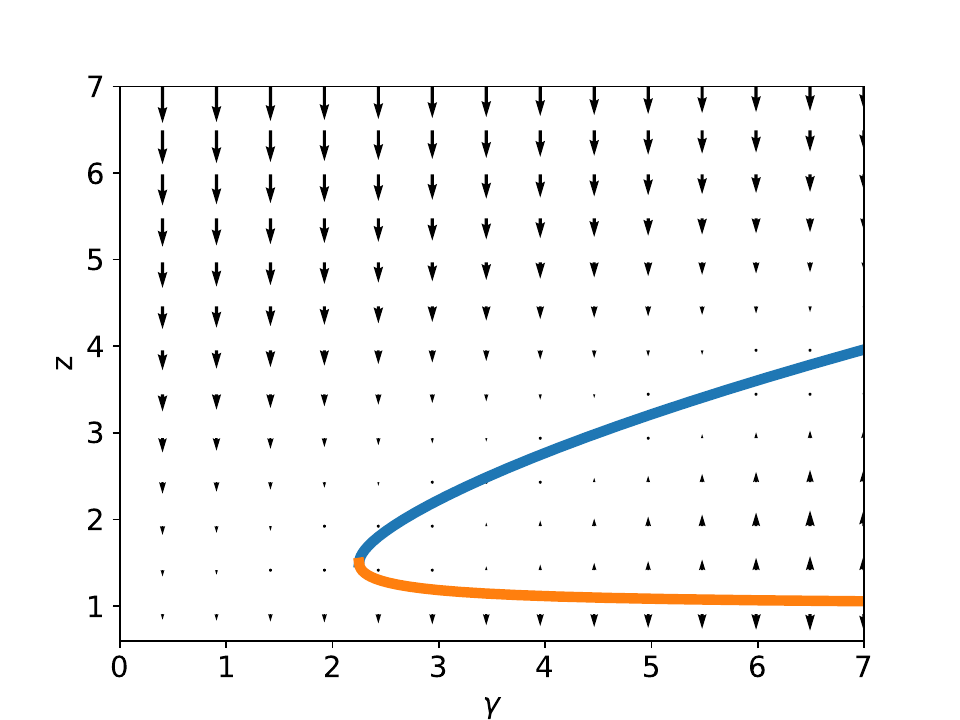}
    \caption{Flow diagram of the rescaled nucleus size $z$ as governed by Eq.~\ref{seq_of_motion_z_3D}. Arrows indicate the direction of $v(z, \gamma)$ for different values of $\gamma$. The orange line denotes the unstable (repelling) fixed point $z_-$, while the blue line marks the stable (attracting) fixed point $z_+$.}
    \label{sfig:flow_diag_3D}
\end{figure}

We draw the corresponding flow diagram with $\gamma$ as a free variable in Fig~\ref{sfig:flow_diag_3D}. Similarly to the previously described 2D case, we note that for $\gamma < 9/4$, the function $v(z)$ doesn't have any zeros in the $z > 0$ region, while for $\gamma > 9/4$, it possesses two real zeros corresponding to a repelling $z_-$ and an attracting fixed point $z_+$ (orange and blue lines in Fig.~\ref{sfig:flow_diag_3D}). \\

The time evolution of $\gamma(\xi)$ is described as follows: In the $\gamma < 9/4$ case, a constant plating current $ j = 3\int_0^\infty \phi(z) (z -1) dz \geq 0$ can not be maintained since

\[3\int_0^\infty \phi_0(z_0) z(z_0, \xi) d z_0 < 3\int_{-v_{max} \xi}^\infty \phi_0(z_0) z_0 d z_0\to 0,\]
where $v_{max} < 0$ is the maximum of $v(z)$. The line of  argumentation closely follows the 2D case, and defines an upper bound for the integrand as $z(z_0, \xi) < z_0$, while the lower integration bound accounts for the minimal initial radius $z_0 > -v_{max} \xi$ that will be dissolved after time $\xi$. Showing that $3 \int \phi(z) z dz \to 0$ then directly leads to $j \leq 0$ in the large $\xi$ limit, such that conversely, perserving the constant current condition forces $\gamma$ to increase to values $\geq 9/4$. \\
In the $\gamma > 9/4$ case, any small fraction of nuclei above the orange line $z_-$ will converge to the blue line $z_+$. Contrary to the 2D discussion, this doesn't lead to a contradiction with the mass conservation but indeed, having all nuclei the same size $z_+$ growing $\propto \tau^{1/3}$ exactly fulfills the constant current condition. We hence summarize that values $\gamma < 9/4$ will increase with progressing time $\xi$, while all $\gamma \geq 9/4$ are expected to be physical asymptotes, with their respective asymptotic distribution being derived in the following paragraph. \\

To obtain the asymptotic distribution we follow the 2D derivation until Eq. \ref{seq_f_cont_z_timeind}
\begin{align*}
    v\frac{\partial \Phi}{\partial z} + \Phi \frac{\partial v}{\partial z} = \alpha \Phi 
\end{align*} 

The corresponding constant current constraint 
\begin{equation*}
    j = 3\int_0^\infty (z - 1) \phi(z) dz = 3 e^{-\alpha\xi} \int_0^\infty (z - 1) \Phi(z) dz
\end{equation*}

leads to $\alpha = 0$. As a side note, we mention here that for the 3D Ostwald ripening with constant volume $\alpha =1$, such that we can obtain the asymptotic distribution by following the 2D discussion verbatim. \\

For our constant current case 

\begin{align}
\label{seq_f_cont_z_timeind3D}
\frac{\partial\Phi(z)v(z)}{\partial z} = 0
\end{align}
and direct integration leads to

\[ \Phi(z)v(z) = c\]
with some integration constant $c$. Interestingly $\Phi(z) = \frac{c}{v(z)}$ does not yield a physical result, as $\frac{1}{v(z)}$ is not normalizable due to divergences at the zeros of $v(z)$. Instead, the physical asymptotic distribution is a more subtle solution of Eq. \ref{seq_f_cont_z_timeind3D} in form of a delta function centered at the attractive fixed point $z_+$

\begin{equation*}
\Phi(z) = A \delta(z - z_+), 
\end{equation*} 
where $A$ is a normalization constant. We discard the solution $\Phi(z) = \delta(z - z_-)$, because $z_-$ is a repelling fixed point in the flow diagram (Fig. \ref{sfig:flow_diag_3D}). \\

We conclude by deriving the following quantities: 
\begin{itemize}
    \item \textbf{Nucleation density}:
\[ \nu = \int A \delta(z-z_+) dz = A\]
\end{itemize}

Unlike in the 2D constant-current ripening case, the nucleation density $\nu$ in 3D remains constant over time and can be treated as an independent variable. The normalization condition, based on mass conservation (Eq.~\ref{seq_mass_cons_3D}), yields:
\[
j=3\int (z-1) A \delta(z- z_+)dz=3\nu(z_+-1)
\]
which gives
\[
\gamma=\frac{\nu}{j}\left(\frac{j}{3\nu}+1\right)^3,
\]
and trivial
\begin{itemize}
    \item \textbf{Nucleus radius}:
    \[
    \rho=\sqrt[3]{\frac{j\tau}{\nu}}
    \]

\end{itemize}

We note here that the 3D constant current case strongly deviates from the constant volume 3D Ostwald ripening, which again sees $\gamma = 9/4$ as the sole possible asymptote. Furthermore, while the constant current case with $\alpha = 0$ leads to a $\delta$-function the constant volume case with $\alpha = 1$ leads to a non-trivial asymptotic distribution with finite width. \\

To conclude, we highlight some qualitative differences between the 2D and 3D ripening behaviors that offer further physical intuition. At the level of the original variables $(\rho, \tau)$, the growth velocity $v_\rho$ in 2D increases monotonically with $\rho$, meaning that even small perturbations to a $\delta$-like distribution cause it to break apart as smaller nuclei dissolve faster. In contrast, the 3D case features a non-monotonic $v_\rho$ with a maximum at finite $\rho$, making the $\delta$-distribution a stable solution: perturbations tend to self-correct, and the system naturally relaxes back to a narrow distribution.

In the rescaled coordinates $(z, \xi)$, this difference manifests in the flow diagram: the unstable $\delta$-solution in 2D is hidden in the $\gamma \to \infty$ limit below the repelling orange line, whereas the stable $\delta$-solution in 3D emerges at $\gamma \geq 9/4$ and corresponds to the attractive fixed point (blue line).

Finally, we comment on why the constant-volume Ostwald ripening result appears similarly to the constant current case in 2D, but not in 3D. In 2D, the instability of the $\delta$-distribution leads both constant-volume and constant-current formulations to converge toward the self-similar asymptotic form. In 3D, however, the physical mechanism underlying volume conservation requires $\rho_s$ to track the distribution center, while under constant current $\rho_s$ can lag behind—effectively decoupling the evolution from constant volume ripening and producing qualitatively different dynamics.

\section{Arrhenius fit of $R_{SEI}$ extracted from Ref~\cite{Wang2019}}
\label{sec_exp_fit}
\begin{table}[h]
    \centering
    \caption{Data extracted from Ref~\cite{Wang2019}. 1M LiTFSI, DOL/DME/LiNO$_3$. Current density 0.25 mA cm$^{-2}$, capacity 0.15 mAh cm$^{-2}$. $v N$ calculated as total capacity devided by $r^3$}
    \begin{tabular}{c|c|c|c}
    $T$ ($^\circ$C)&$R_{SEI}$ ($ V_m$)&$r$ ($\mu$m)$^b$&$v N$  ($\mu$m$^{-2}$)\\
    \hline
    -20 &224	&0.36	&15.5\\
    -10 &157$^*$	&0.63	&2.9\\
    0   &134	&1.1	&0.63\\
    10  &56$^*$ 	&1.4	&0.28\\
    20	&32 	&1.7	&0.15\\
    30	&23$^*$ 	&2.2	&0.067\\
    \end{tabular}\\
    \footnotesize{$^*$ calculated from linear regression in Arrhenius coordinates $\ln(R)\sim-T^{-1}$}\\
\end{table}
\begin{figure}[!h]
    \centering
    \includegraphics[width=0.25\linewidth]{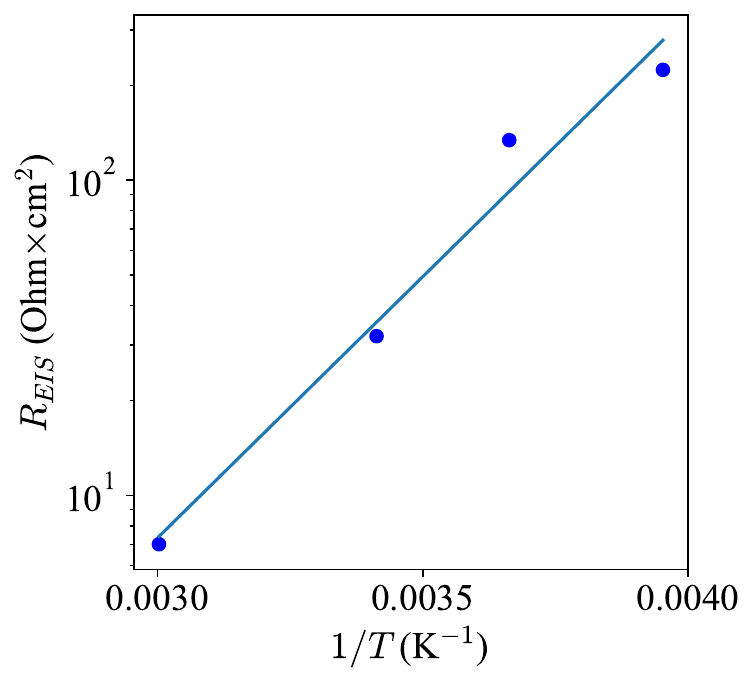}
    \caption{Linear regression of $R_{SEI}$ in Arrhenius coordinates $\ln(R)\sim-T^{-1}$}
    \label{fig:R_ct_fit}
\end{figure}

\printbibliography[keyword={SI}]

\end{document}